# Nonreciprocal surface acoustic wave propagation via magneto-rotation coupling

**Authors**

Mingran Xu,[1,2] Kei Yamamoto,[3,2] Jorge Puebla,[2,*] Korbinian Baumgaertl,[4] Bivas Rana,[2] Katsuya Miura,[5] Hiromasa Takahashi,[5] Dirk Grundler,[4,6] Sadamichi Maekawa,[2,3,7] and Yoshichika Otani[1,2,8,*]

**Affiliations**

[1]Institute for Solid State Physics, University of Tokyo, 5-1-5 Kashiwanoha, Kashiwa, Chiba, 277-8581, Japan

[2]CEMS, RIKEN, Saitama, 351-0198, Japan

[3]Advanced Science Research Center, Japan Atomic Energy Agency, Tokai 319-1195, Japan

[4]Laboratory of Nanoscale Magnetic Materials and Magnonics(LMGN), Institute of Materials (IMX), School of Engineering, Ecole Polytechnique Fédérale de Lausanne (EPFL), 1015 Lausanne, Switzerland

[5]Research and Development Group, Hitachi, Ltd., 1-280 Higashi-koigakubo, Kokubunji-shi, Tokyo 185-8601, Japan

[6]Institute of Microengineering (IMT), School of Engineering, Ecole Polytechnique Fédérale de Lausanne (EPFL), 1015 Lausanne, Switzerland

[7]Kavli Institute for Theoretical Sciences, University of Chinese Academy of Sciences, Beijing 100049, People's Republic of China

[8]CREST, Japan Science and Technology Agency, Kawaguchi, Saitama 332-0012, Japan

∗ Corresponding author: jorgeluis.pueblanunez@riken.jp, yotani@riken.jp

**Abstract**

One of the most fundamental forms of magnon-phonon interaction is an intrinsic property of magnetic materials, the "magnetoelastic coupling". This particular form of interaction has been the basis for describing magnetic materials and their strain related applications, where strain induces changes of internal magnetic fields. Different from the magnetoelastic coupling, more than 40 years ago, it was proposed that surface acoustic waves may induce surface magnons via rotational motion of the lattice in anisotropic magnets. However, a signature of this magnon-phonon coupling mechanism, termed magneto-rotation coupling, has been elusive. Here, we report the first observation and theoretical framework of the magneto-rotation coupling in a perpendicularly anisotropic ultra-thin film Ta/CoFeB(1.6 nm)/MgO, which consequently induces nonreciprocal acoustic wave attenuation with an unprecedented ratio up to 100% rectification at the theoretically predicted optimized condition. Our work not only experimentally demonstrates a fundamentally new path for investigating magnon-phonon coupling, but also justify the feasibility of the magneto-rotation coupling based application.

**Introduction**

In a general description, a rectification consists of passing signals in one direction while suppressing those in the opposite direction in a counter-propagation scenario. The best-known example of a rectifier is the electronic diode which converts alternating current (AC) to direct current (DC), allowing the development of the huge electronic industry we have today. Despite the great success of the electronic rectifier, challenges remain open, such as efficient rectifiers of small dimensions at high frequencies. Therefore, rectification of other energy entities has been intensively explored, in the form of acoustic rectifiers(*1, 2*), thermal rectifiers(*3*), magnon rectifiers(*4*), and photon rectifiers(*5*). Here we demonstrate a giant nonreciprocal behavior of an on-chip acousto-magnetic rectifier at room temperature and GHz frequency. Our device exploits the magnon-phonon coupling by which surface acoustic waves (SAWs) interact with ferromagnetic (FM) films and consequently generate spin waves (SWs).

At a resonance condition, the coupling of SAWs with magnetic films produces acoustically driven ferromagnetic resonance (a-FMR)(*6, 7*). Consequently, the a-FMR generates a spin current which can be converted to charge current by the inverse

Edelstein effect (IEE)(*8*) or spin Hall effect (ISHE)(*9*). Evidence of nonreciprocal behaviors in attenuation of amplitudes was reported when SAWs interacted with magnetic films(*8, 10, 11*). In these works, the origin of the nonreciprocal behaviors was attributed to magneto-elastic coupling induced attenuation or the interference between the longitudinal and shear components of the strain tensor in SAWs. However, in the thin film limit, $kd \ll 1$, where $k > 0$ is the absolute value of the wavenumber of SAWs and $d$ the film thickness, the shear strain is strongly diminished and the remaining longitudinal strain is expected to induce only reciprocal magnetization dynamics, ergo limiting the nonreciprocity that can be achieved. Interestingly, for Rayleigh type SAWs, there is an additional dynamical component that survives in the thin film limit, the rotation tensor of elastic deformation, $\omega_{ij} = \frac{1}{2}(\frac{\partial u_i}{\partial x_j} - \frac{\partial u_j}{\partial x_i})$, which describes the rotational deformation of the lattice. Here, $u_i, i = x, y, z$ are Cartesian components of the elastic deformation vector field. The nonvanishing $\omega_{ij}$ implies that the individual lattice points undergo a rotation per wave cycle, whose chirality changes its sign according to the wave propagation direction (see the blue and red oriented circles in Fig. 1). Furthermore, this rotational term can also couple to the magnetization via magnetic anisotropy, according to the theoretical prediction by Maekawa and Tachiki (*12*). Therefore, extending the previous model, we take account of *magneto-rotation coupling*, which turns out to be crucial for the giant nonreciprocity in the present work. The rectification effect that we observe is far larger than the record values of 20% (recently reported in Ref (*11*)). Our findings go beyond their work in that we consider theoretically the magneto-rotation coupling and thereby explain the experimental value of 100% at optimized experimental conditions. This intriguing prediction is in contrast to Ref (*11*) where it was speculated that the nonreciprocal attenuation depended critically on the magnetic damping.

**Results**

Fig. 2A shows the schematics of SAWs propagation through a heterostructure, which consists of four layers, Ta(10 nm)/Co$_{20}$Fe$_{60}$B$_{20}$(1.6 nm)/MgO(2 nm)/Al$_2$O$_3$(10 nm), son a piezoelectric substrate, $Y$ cut LiNbO$_3$. The acoustic waves are excited by applying an rf voltage on the input port of interdigital transducers (IDTs), which were patterned by electron beam lithography. Due to the inverse piezoelectric effect, the rf voltage at a frequency of 6.1 GHz induces vibrations of the lattice, launching SAWs propagating parallel to the $x$-axis (see the coordinate system in Fig. 2A). When SAWs propagate through the heterostructure, the oscillation of the lattice points inside the magnet induces an effective rf magnetic field via cubic magnetoelastic coupling and the magneto-rotation coupling. Under a static in-plane external magnetic field $\mathbf{H} = H(\cos\phi, \sin\phi, 0)$, spin waves are excited, which results in SAW attenuation (see Fig. 2B). After passing through the heterostructure, the remaining SAWs are converted back into an rf voltage signal via piezoelectric effect on the output port IDTs. The attenuation was characterized by a vector network analyzer (VNA), based on the scattering parameters, S21 and S12. By measuring S21 and S12, we investigated the SAWs propagating along $+x$ and $-x$ direction, respectively, referred to as $+k$ and $-k$ from here on.

Fig. 2B presents attenuation spectra for SAW($+k$) and SAW($-k$) when an external magnetic field was applied at $\phi = 10°$ in the $xy$ plane. The external magnetic field was initially set to 200 mT to saturate the magnetic film and then swept from 120 to 70 mT in 0.5 mT steps. Acoustic ferromagnetic resonance is obtained at an external field value of 96 mT, inducing magnetic field dependent SAW attenuations. However, in the spectra, the SAW($+k$) shows a negligible attenuation $P_{+k}$ while SAW($-k$) shows a relatively large attenuation $P_{-k}$. The large difference indicates a strong nonreciprocal behavior, and therefore a strong rectifier effect on acoustic waves. From this, we extract the NR ratio $(P_{+k} - P_{-k})/(P_{+k} + P_{-k})$, which depends on the magnetic field direction, and plot it in Fig. 3. The NR shows a strong dependence on the magnetic field direction with respect to the SAW propagation direction $\hat{\mathbf{x}}$, reaching a maximum value of 100%.

To understand the origin of the giant nonreciprocity, we theoretically model the magnetization dynamics driven by propagating SAWs in FM thin films. Treating the film as an isotropic elastic body, SAWs propagating along *x*-axis are fully characterized by the non-vanishing components $\varepsilon_{xx}$, $\varepsilon_{xz}$, $\varepsilon_{zz}$ of the strain tensor $\varepsilon_{ij} = \frac{1}{2}(\frac{\partial u_i}{\partial x_j} + \frac{\partial u_j}{\partial x_i})$ and the rotation $\omega_{xz}$. Among them, the shear strain $\varepsilon_{xz}$ scales as $kd$ when the film thickness $d$ is small and becomes negligible compared to the others in our setup where $kd < 10^{-2}$. Thus the mechanism (*8, 10, 11*) proposed in cannot account for the pronounced nonreciprocity presented above and another explanation is required. As predicted in (*12, 13*), a perpendicular magnetic anisotropy, present in our heterostructure, enables the rotation $\omega_{xz}$ to generate an effective magnetic field acting on the magnetization. Combined with the conventional magnetoelastic effect that results in an additional effective field proportional to the strain tensor $\varepsilon_{ij}$, the total effective field projected onto the plane perpendicular to the normalized ground state magnetization n yields h = $h_\perp \hat{z} + h_\parallel \hat{v}$ where $\hat{v} = \hat{z} \times$ n and

$$h_\perp = \frac{\gamma}{\mu_0 M_s}(2K_u\omega_{xz} - b_2\varepsilon_{xz})\cos\phi, \quad h_\parallel = \frac{\gamma b_1}{\mu_0 M_s}\varepsilon_{xx}\sin2\phi \quad (1)$$

with the cubic magnetoelastic coefficients $b_{1,2}$, the perpendicular uniaxial anisotropy $K_u$, the gyromagnetic ratio $\gamma < 0$, the saturation magnetization $M_s$ and the permeability of vacuum $\mu_0$. Further noting that $\omega_{xz}$ has exactly the same phase and $k$ dependence as those of $\varepsilon_{xz}$, therefore, we conclude that magneto-rotation coupling is capable of replacing the shear strain as a source of nonreciprocal SAW attenuation. In addition, $\omega_{xz}$ is greater by a factor of $(kd)^{-1}$ than $\varepsilon_{xz}$ in the thin film limit so that the resulting nonreciprocity tends to be much larger when the magnetic anisotropy is comparable to the magnetostriction, which is the case for CoFeB.

The SAW attenuation $P_{\pm k}$ is related to the power dissipated by the spin waves excited by the elastic effective field **h**. It can be readily computed as a function of the magnetization angle $\phi$, whose details are given in the supplemental material. The formula taking into account of exchange interaction, the uniaxial and cubic magnetic anisotropy, dipole-dipole interaction, and the Dzyaloshinskii-Moriya interaction (DMI) induced by the inversion symmetry breaking at the interface has been used to fit the data in Fig. 3. The agreement is quantitative, suggesting that the magneto-rotation coupling is indeed responsible for the giant nonreciprocity.

So far, we focused on the nonreciprocal behavior of SAW attenuation $P_{\pm k}$. In addition, we observed a nonreciprocal behavior in the resonance field $H^{res}_{\pm k}$ when the SAW wavenumber is reversed from $k$ to $-k$. The DMI is the antisymmetric exchange coupling between neighboring magnetic spins, which gives a contribution to the local energy that is linear in spatial derivatives. Therefore, it leads to an asymmetry in spin wave dispersion relation with respect to the sign of wavenumber $\pm k$. Commonly, the strength of the DMI or DMI coefficient $D$ is determined by measuring the difference in resonance frequencies between two spin waves propagating in opposite directions (opposite wavenumbers $\pm k$), which is very often achieved by Brillouin light scattering spectroscopy (BLS) (*14*). However, as the magnetic layers become thinner, the magnetic response becomes weaker, which consequently challenges the precise measurement of the DMI coefficient.

Interestingly with the a-FMR, we observe a clear difference in the resonance condition of the acoustic waves with the spin waves in the 1.6 nm thick CoFeB layer (see Fig. 4A). After obtaining the resonance field $H^{res}_{\pm k}(\phi)$ as a function of the angle $\phi$ for $\pm k$ through Lorentzian fits of line shapes (Fig. 4A), we estimate $D$ by fitting the angular dependence of the resonance field difference $\Delta H^{res}(\phi) \equiv H^{res}_{+k}(\phi) - H^{res}_{-k}(\phi)$ by

$$\Delta H^{\text{res}}(\phi) = \frac{8D\omega k \sin\phi}{|\gamma|\mu_0^2 M_s \sqrt{(H_v - H_z)^2 + 4(\omega/\gamma\mu_0)^2}} \quad (2)$$

where $H_v - H_z$ depends only on the saturation magnetization, the anisotropy constants, $k$ and $\cos^2\phi$ (see supplemental material). As can be seen in Fig. 4B, the observed $\Delta H^{\text{res}}(\phi)$ follows the $\sin\phi$ dependence expected for the DMI, yielding $D = 0.089 \pm 0.011\, mJ/m^2$, in good agreement with previous reports (*15*) and our BLS characterization (see supplemental material). This suggests that the acoustic FMR may also serve as an efficient and accurate means of determining the DMI coefficient in magnetic thin films. In addition, for commercial BLS, the maximum wavenumber $k$ that can be explored is limited by the wavelength of the laser, $k_{\max}(\text{BLS}) = \frac{2\pi}{\lambda_{\text{laser}}/2} \approx 3 \cdot 10^7\, \text{rad/m}$, where $\lambda$ is in the visible range. In contrast, $k$ of the SAW, which is determined by the pitch resolution of electron beam lithography (EBL), $k_{\max}(\text{SAW}) = \frac{2\pi}{4\lambda_{\text{EBL}}}$, where $\lambda_{\text{EBL}} \approx 10$ nm (*16*), is capable of reaching $10^8$ rad/m level.

We note that the obtained value of *D* is too small to account for the nonreciprocity in the SAW attenuation by the mechanism proposed in (*17*).

**Discussion**

In conclusion, we demonstrated strongly nonreciprocal acoustic attenuation in power (Fig. 2B and resonance field (Fig. 4A), separately. These intriguing nonreciprocal features of the presented acoustic devices suggest an extraordinary versatility of acousto-magnetic applications. The drastic angular dependence of the nonreciprocal ratio (Fig. 3) indicates an efficient and adjustable acoustic rectifier, and the systematic change of nonreciprocity in resonance field (Fig. 4B) presents its capability as a new route for characterization of DMI. Besides, since the nonreciprocity introduced here stems from the magnetic anisotropy, it can be further modulated by external electric field, as has been reported for the CoFeB/MgO interface (*18*). Considering the wide application of the general acoustic device in sensing, filtering and information transportation, utilization of acoustic-magneto rectifier would not only provide highly accurate methods for sensing magnetic properties, but also further advance the present acoustic technology, and eventually push the development of acousto-magnetic logic devices as an attractive alternative to their magnonic counterparts(*19*, *20*).

**Materials and Methods**
**Device fabrication**
Fig. 2A shows the schematics of SAWs propagation through a heterostructure, which consists of four layers, Ta(10 nm)/$Co_{20}Fe_{60}B_{20}$(1.6 nm)/MgO(2 nm)/$Al_2O_3$(10 nm), grown by radio-frequency (rf) sputtering on a piezoelectric substrate, *Y* cut $LiNbO_3$.

**Acknowledgments**

**General**: We would like to thank V.S. Bhat, M. Andrea and S. Watanabe for their kind help in the initial stage of the experiment, O. Gomonay and K. Sato for helpful comments, M. Ishida for her support in preparing figures, and Emergent Matter Science Research Support Team in RIKEN for the technical support, and Dongshi Zhang for reviewing the paper.

**Funding:** This work was supported by Grants-in-Aid for Scientific Research on Innovative Areas (No. 26103001, No. 26103002) and JSPS KAKENHI (No. 19H05629). MX would like to thank support from JSPS through "Research program for Young Scientists" (No. 19J21720) and RIKEN IPA Program. KY would like to acknowledge support from JSPS KAKENHI (JP 19K21040) and the Inter-university Cooperative Research Program of the Institute for Materials Research, Tohoku University (Proposal No. 19K0007). SM was financially supported by ERATO, JST, and KAKENHI (No 17H02927 and No. 26103006) from MEXT, Japan. K.B. and D.G. thank SNSF for funding via Grant No. 163016. Also, this work was partially supported by CREST(JPMJCR18T3), Japan Science and Technology Agency (JST).

**Author contributions:** M.X. fabricated the samples and performed transmission measurement. K.Y. formulated the theoretical model. M.X. and K.Y. analyzed the data. K.Y. and S.M. developed the explanation of the experimental results. K.B. performed the BLS measurement. K.M. and H.T grew the CoFeB film. M. X., K.Y, J.P., K.B., B.R., K.M., H.T., D. G., S.M., and Y. O discussed the results and commented on the manuscript. J.P., S.M. and Y.O. supervised the project.

**Competing interests:** The authors declare no competing financial interests.

**Data and materials availability:** Both data and materials are available under reasonable request to corresponding authors.

**Figures and Tables**

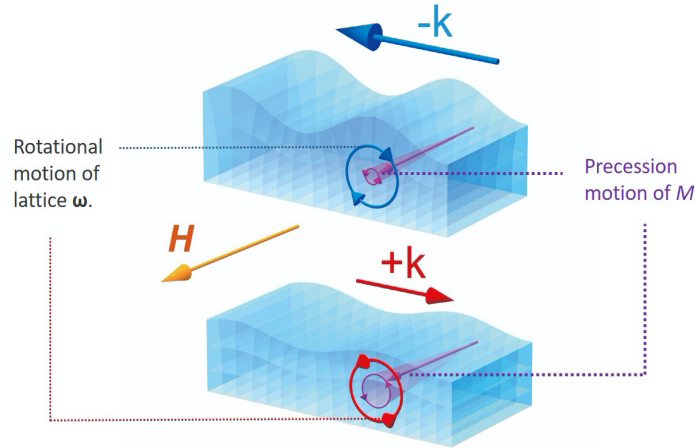

**Fig. 1. Schematics of the magneto-rotation coupling.** Depending on the propagation direction, surface acoustic waves rotate the lattice in opposite directions (as indicted by the blue and red oriented cycles in the figure). This rotational motion couples with the magnetization via magnetic anisotropies, giving rise to a circularly polarized effective field, which either suppresses or enhances the magnetization precession (purple cone), and in turn induces a nonreciprocal attenuation on the surface acoustic waves.

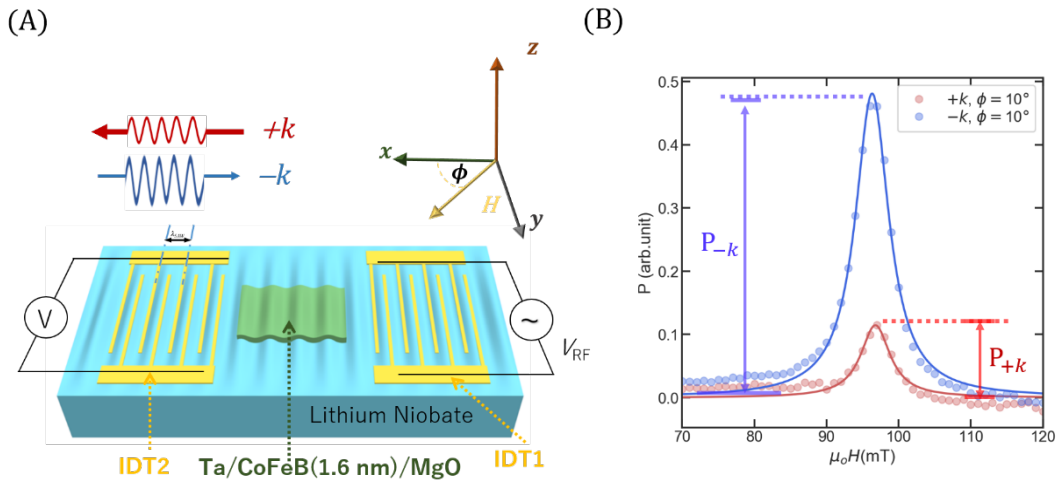

**Fig. 2. Nonreciprocal propagation of acousto-magnetic waves in Ta/CoFeB/MgO.** **(A),** Device schematics of surface acoustic waves coupling to a ferromagnetic layer at GHz frequencies. **(B),** Attenuation of acoustic waves, $P_{\pm k}$, near a spin-wave resonance condition for surface acoustic wavenumbers $+k$ and $-k$.

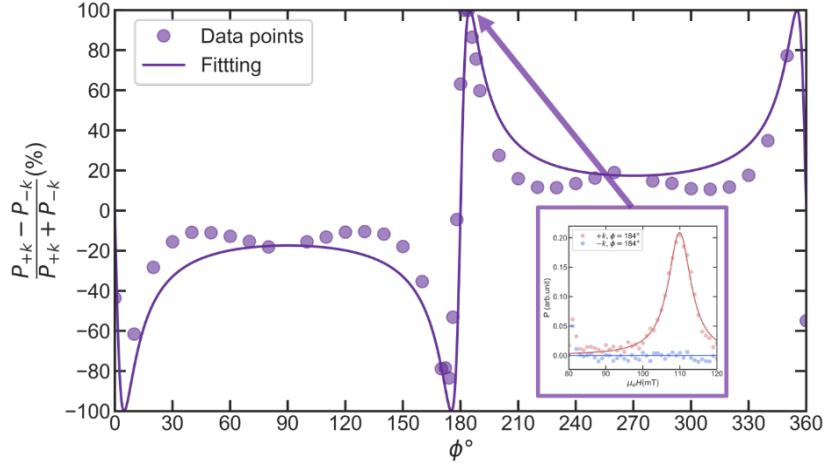

**Fig. 3. Nonreciprocal ratio dependence on magnetic field direction**. The magnetic field direction is varied with respect to the SAW propagation direction $\hat{x}$. We observe a strong variation of the nonreciprocal ratio, reaching a maximum value of 100% at $\phi = 184°$ (see the insert).

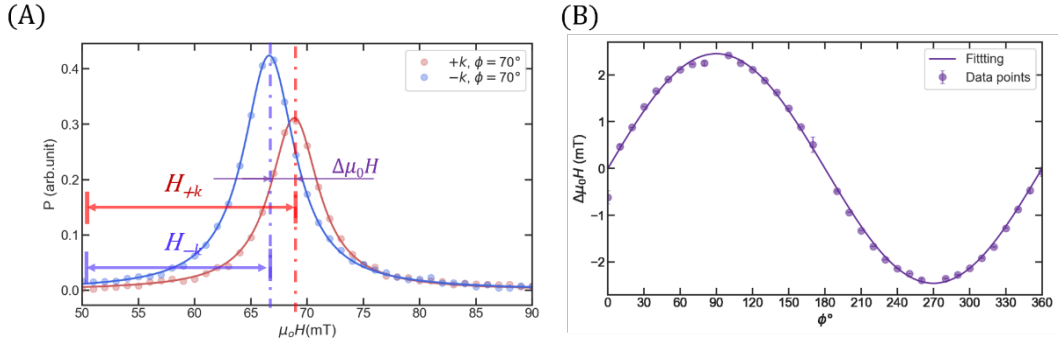

**Fig. 4. Assessment of Dzyaloshinskii Moriya Interaction by nonreciprocal magnon-phonon interaction**. **(A)**, Resonance field difference $\Delta H_{res}$ between acousto-magnetic waves induced by surface acoustic waves propagating in $+k$ and $-k$. **(B)**, An gle dependence of the resonance field difference between acousto-magnetic waves induced by surface acoustic waves with wavenumbers $+k$ and $-k$ fitted according to Eq. (2).

# Supplementary Materials

**Section S1. Theoretical modelling**

We present a theoretical description of surface acoustic wave absorptions by a ferromagnetic thin film. We model the experimental setup by an isotropic elastic material filling the half-space $z < d/2$ with its top layer of thickness $d$ assumed to be ferromagnetic with magnetization $\boldsymbol{M}$. Although the elastic properties could be anisotropic and differ between the magnet and the substrate, this simplified model turns out to be sufficient for our purposes. Let $u_{x,y,z}$ be the Cartesian components of the displacement vector field. The acoustic waves in isotropic media are characterized by Lamé constants $\lambda, \mu$ through the elastic free energy

$$F = \int d^3r \left\{ \frac{\lambda}{2}\left(\varepsilon_{xx} + \varepsilon_{yy} + \varepsilon_{zz}\right)^2 + \mu\left(\varepsilon_{xx}^2 + \varepsilon_{yy}^2 + \varepsilon_{zz}^2 + 2\varepsilon_{xy} + 2\varepsilon_{yz} + 2\varepsilon_{zx}\right) \right\} \quad (S1)$$

where $\varepsilon_{ij} = (\partial_j u_i + \partial_i u_j)/2$ are the components of the strain tensor. The dynamics of $\boldsymbol{u}$ with the isotropic free energy is studied in any textbook on continuum mechanics. In particular, the longitudinal and transverse bulk acoustic waves travel at respective speeds of sound $c_L = \sqrt{(\lambda + 2\mu)/\rho}$, $c_T = \sqrt{\mu/\rho}$ where $\rho$ is the mass density, and the surface acoustic waves propagating in positive and negative $x$ directions with respective wavenumbers $\pm k, k > 0$ along the surface $z = d/2$ are described by the solution

$$\begin{pmatrix} u_x^\pm \\ u_z^\pm \end{pmatrix} = C \mathrm{Re}\left[\begin{pmatrix} (1 - \xi_T^2/2)\{2e^{\kappa_L(z-d/2)} - (2-\xi_T^2)e^{\kappa_T(z-d/2)}\} \\ \mp i\sqrt{1-\xi_L^2}\{(2-\xi_T^2)e^{\kappa_L(z-d/2)} - 2e^{\kappa_T(z-d/2)}\} \end{pmatrix} e^{-i(\omega t \mp kx)}\right] \quad (S2)$$

where $C$ is an arbitrary real constant, and the parameters $\kappa_{L,T}, \xi_{L,T}$ are related to the bulk speeds of sound $c_{L,T}$ and the surface speed of sound $c_S$ by

$$\kappa_{L,T} = \sqrt{k^2 - \frac{\omega^2}{c_{L,T}^2}}, \xi_{L,T} = \frac{c_S}{c_{L,T}}. \quad (S3)$$

Note that the value of $c_S$ depends implicitly on $c_{L,T}$ through the algebraic equation

$$\xi_T^6 - 8\xi_T^4 + 8\left(3 - \frac{2c_T^2}{c_L^2}\right)\xi_T^2 - 16\left(1 - \frac{c_T^2}{c_L^2}\right) = 0. \quad (S4)$$

The "spin"-momentum locking of surface acoustic waves manifests itself in the phase difference between $u_x^\pm$ and $u_z^\pm$ by $\mp i = e^{\mp i\pi/2}$, which changes the sign under $k \to -k$.

We regard the surface acoustic wave solution (S2) with fixed $\omega$ and $k$ being given as an effective rf field and study how the magnet responds. Back reactions of magnetization dynamics onto acoustic waves are neglected. The purely magnetic part of the free energy is assumed to include an external magnetic field, cubic crystalline and interface-induced uniaxial magnetic anisotropies, and exchange, dipole-dipole and Dzyaloshinskii-Moriya interactions:

$$W = \int d^3r [-\mu_0 M_s \boldsymbol{H} \cdot \boldsymbol{n} - \frac{A}{2} \boldsymbol{n} \cdot \nabla^2 \boldsymbol{n} + \frac{\mu_0 M_s^2}{8\pi} \int d^3 r' (\boldsymbol{n} \cdot \nabla)(\boldsymbol{n}' \cdot \nabla') \frac{1}{|\boldsymbol{r} - \boldsymbol{r}'|}$$
$$+ K_c(n_x^2 n_y^2 + n_y^2 n_z^2 + n_z^2 n_x^2) - K_\perp n_z^2 - K_\parallel n_x^2 + D\boldsymbol{n}$$
$$\cdot \{(\hat{\boldsymbol{x}}\partial_y - \hat{\boldsymbol{y}}\partial_x) \times \boldsymbol{n}\}], (S5)$$

where $M_s$ is the saturation magnetization, $\mathbf{n} = \mathbf{M}/M_s$, $\mathbf{n}'$ is the value of $\mathbf{n}$ evaluated at $\mathbf{r}'$, $\nabla'$ is the spatial derivative with respect to $\mathbf{r}'$ and $\hat{\mathbf{x}}, \hat{\mathbf{y}}$ are unit vectors in $x$ and $y$ directions. Note that $K_\perp$ arises from the interface inversion symmetry breaking while $K_\parallel$ is present due to the crystallographic $c$-axis alignment of the single crystalline LiNbO$_3$ substrate. Taking the external field to be spatially homogeneous and in-plane $\mathbf{H} = H(\cos\phi, \sin\phi, 0)$, the ground state magnetization is also in the plane $\mathbf{n} = (\cos\theta, \sin\theta, 0)$ and $\theta = \phi$ for sufficiently strong $\mathbf{H}$ if $K_c = K_\parallel = 0$. The spin waves excited by the surface acoustic waves $\mathbf{u}^\pm$ have wavevectors $\pm k\hat{\mathbf{x}}$ respectively. In the thin film limit $kd \ll 1$, linear perturbation around the ground state yields the dispersion relation

$$\omega_k = |\gamma|\mu_0 \sqrt{\{H\cos(\theta - \phi) + H_v(\theta)\}\{H\cos(\theta - \phi) + H_z(\theta)\}} + \gamma\mu_0 H_{\text{DMI}}(\theta), (S6)$$

where $|\gamma|$ is the gyromagnetic ratio and

$$H_v(\theta) = \frac{Ak^2}{\mu_0 M_s} + M_s\left(1 - \frac{1-e^{-kd}}{kd}\right)\sin^2\theta + \frac{2K_\parallel \cos 2\theta - K_c(1 - 3\cos^2 2\theta)}{\mu_0 M_s} \quad (S7)$$

$$H_z(\theta) = \frac{Ak^2}{\mu_0 M_s} + M_s \frac{1-e^{-kd}}{kd} - \frac{2K_\perp - 2K_\parallel \cos^2\theta - K_c(1+\cos^2 2\theta)}{\mu_0 M_s} \quad (S8)$$

$$H_{DMI}(\theta) = \pm \frac{2Dk\sin\theta}{\mu_0 M_s} \quad (S9)$$

Note that we throughout use the convention to take frequencies to be positive. In the main text, we chose not to discuss $K_\parallel$ and $K_c$, which we shall see are small, and denoted the total perpendicular anisotropy by $K_u = K_\perp - \mu_0 M_s^2/2$. For a given driving frequency $\omega$, the resonance field is determined by

$$H^{\text{res}} \cos(\theta - \phi) = \frac{-(H_v + H_z) + \sqrt{(H_v - H_z)^2 + 4(H_{DMI} - \omega/\gamma\mu_0)^2}}{2}. \quad (S10)$$

and expanding the square root to linear order in $H_{DMI}$ (and setting $\theta = \phi$) yields Eq. (2) in the main text.

There are a variety of ways in which **M** interacts with acoustic waves. For cubic crystals, one usually includes the conventional magnetoelastic coupling in the free energy

$$I_1 = \int d^3r \{b_1(n_x^2 \varepsilon_{xx} + n_y^2 \varepsilon_{yy} + n_z^2 \varepsilon_{zz}) + 2b_2(n_x n_y \varepsilon_{xy} + n_y n_z \varepsilon_{zx} + n_z n_x \varepsilon_{zx})\}. \quad (S11)$$

It turns out to be insufficient, however, for explaining the large non-reciprocal response seen in our experiment. For this purpose, we consider free energy terms that describe interactions between **M** and the rotation of elastic deformations $\omega_{ij} = (\partial_j u_i - \partial_i u_j)/2$. There are many possible mechanisms for magneto-rotation coupling, and here we discuss those that are directly related to the purely magnetic free energy $W$. As first pointed out by Maekawa and Tachiki(*12*), magnetic anisotropy fields induce magneto-rotation couplings through reorientations of crystalline axes, which for uniaxial and cubic anisotropies read

$$I_2 = 2K_\perp \int d^3r (\omega_{zx} n_x + \omega_{zy} n_y) n_z + 2K_\parallel \int d^3r (\omega_{xy} n_y + \omega_{xz} n_z) n_x$$
$$+ 2K_c \int d^3r \{n_x n_u (n_x^2 - n_y^2) \omega_{xy} + n_y n_z (n_y^2 - n_z^2) \omega_{yz}$$
$$+ n_z n_x (n_z^2 - n_x^2) \omega_{zx}\}, \quad (S12)$$

Similarly, the dipolar shape anisotropy results in a magneto-rotation coupling via change of the surface normal directions induced by SAWs. Based on the model of coupling between magnons and surface deformations given in (*21*), one derives the interaction energy

$$I_3 = \frac{\mu_0 M_s^2}{8\pi} \int d^3r \int d^3r' [\{\delta(z - \frac{d}{2}) - \delta(z + \frac{d}{2})\} u_z(\mathbf{r})$$
$$+ \{\delta(z' - \frac{d}{2}) - \delta(z' + \frac{d}{2})\} u_z(\mathbf{r}')](\mathbf{n} \cdot \nabla)(\mathbf{n}' \cdot \nabla') \frac{1}{|\mathbf{r} - \mathbf{r}'|}. \quad (S13)$$

Although this contains both strain and rotation, we shall see shortly that for in-plane magnetization and surface acoustic waves, the strain can be neglected. The Dzyaloshinkii-Moriya interactions are also affected by the crystalline orientation and consequently generate couplings between magnet and lattice deformations;

$$I_4 = D \int d^3r \, \mathbf{n} \cdot [\{\sum_{a=y,z}(\varepsilon_{xa} - \omega_{xa})\hat{\mathbf{r}}_a \partial_y - \sum_{a=x,z}(\varepsilon_{ya} - \omega_{ya})\hat{\mathbf{r}}_a \partial_x + \hat{\mathbf{x}} \sum_{a=z,x}(\varepsilon_{ya} - \omega_{ya})\partial_a$$
$$- \hat{\mathbf{y}} \sum_{a=y,z}(\varepsilon_{xa} - \omega_{xa})\partial_a\} \times \mathbf{n}], \quad (S14)$$

where $\hat{\mathbf{r}}_x = \hat{\mathbf{x}}, \hat{\mathbf{r}}_y = \hat{\mathbf{y}}, \hat{\mathbf{r}}_z = \hat{\mathbf{z}}$ Again, the strain couplings are discarded later. Finally, when the microscopic magnetic moments may be considered fixed on individual atomic sites and adiabatically following the motion of the lattice, there will be an analogue of Coriolis force called spin-rotation coupling (22) given by

$$I_5 = \frac{\hbar S}{V} \int d^3r (n_x \partial_t \omega_{yz} + n_y \partial_t \omega_{zx} + n_z \partial_t \omega_{xy}), \quad (S15)$$

where $S/V$ is the effective length of spin per unit cell.

Each of the interaction terms introduces an effective rf field $\mathbf{h}_a$ acting on the magnetization $\mathbf{M}$ where $\mathbf{h}_a = -\delta I_a / \delta(\mu_0 M_s \mathbf{n}), a = 1, \cdots, 5$. Evaluating these fields for the ground state configuration of $\mathbf{M}$ yields

$$\mathbf{h}_1 = \frac{b_1(\varepsilon_{xx} - \varepsilon_{yy})\sin 2\theta - 2b_2 \epsilon_{xy} \cos 2\theta}{\mu_0 M_s} \hat{\mathbf{v}} - \frac{2b_2(\epsilon_{zx}\cos\theta + \epsilon_{zy}\sin\theta)}{\mu_0 M_s} \hat{\mathbf{z}}, (S16)$$

$$\mathbf{h}_2 = -\frac{2K_{\parallel}\cos 2\theta - 2K_c \cos 4\theta}{\mu_0 M_s} \omega_{xy} \hat{\mathbf{v}}$$
$$- \frac{2K_{\perp}(\omega_{zx}\cos\theta + \omega_{zy}\sin\theta) - 2K_{\parallel}\omega_{zx}\cos\theta - 2K_c(\omega_{zx}\cos^3\theta + \omega_{yz}\sin^3\theta)}{\mu_0 M_s} \hat{\mathbf{z}}, (S17)$$

$$\mathbf{h}_3 = \frac{M_s}{4\pi}\{\hat{\mathbf{v}}(\hat{\mathbf{v}}\cdot\nabla)+\hat{\mathbf{z}}\partial_z\}$$

$$\times \int d^3r' \left\{\delta\left(z'-\frac{d}{2}\right)-\delta\left(z'+\frac{d}{2}\right)\right\}\frac{1}{|\mathbf{r}-\mathbf{r}'|}(\cos\theta\partial_{x'}$$

$$+\sin\theta\partial_{y'})u_z(\mathbf{r}'), \quad (S18)$$

$$\mathbf{h}_4 = \frac{D}{\mu_0 M_s}[\{\partial_y(\varepsilon_{zx}+\omega_{zx})-\partial_x(\varepsilon_{yz}-\omega_{yz})\}\hat{\mathbf{v}}$$

$$+\partial_z\{(\varepsilon_{zx}+\omega_{zx})\cos\theta+(\varepsilon_{yz}-\omega_{yz})\sin\theta\}\hat{\mathbf{z}}], \quad (S19)$$

$$\mathbf{h}_5 = -\frac{\hbar S}{\mu_0 M_s V}\partial_t\{(\omega_{zx}\cos\theta-\omega_{yz}\sin\theta)\hat{\mathbf{v}}+\omega_{xy}\hat{\mathbf{z}}\}, \quad (S20)$$

where $\hat{\mathbf{v}}=-\hat{\mathbf{x}}\sin\theta+\hat{\mathbf{y}}\cos\theta\perp\mathbf{n}_0$ and the in-plane components have been projected onto this axis. To obtain Eq. (1) of the main text, we set $K_\parallel = K_c = 0$ and discarded $\mathbf{h}_4$ and $\mathbf{h}_5$, which is justified in the next section. The main contribution of $\mathbf{h}_3$ is to replace $K_\perp$ in $\mathbf{h}_2$ by $K_u$. For the surface acoustic waves $\mathbf{u}^\pm$ propagating in the positive and negative $x$ directions, their non-vanishing components are given as follows:

$$\varepsilon_{xx}^\pm = \pm iC\left(1-\frac{\xi_T^2}{2}\right)\left\{e^{\kappa_L(z-\frac{d}{2})}-\left(1-\frac{\xi_T^2}{2}\right)e^{\kappa_T(z-\frac{d}{2})}\right\}, (S21)$$

$$\varepsilon_{zz}^\pm = \mp iC\left(1-\frac{\xi_T^2}{2}\right)\left\{(1-\xi_L^2)e^{\kappa_L(z-\frac{d}{2})}-\left(1-\frac{\xi_T^2}{2}\right)e^{\kappa_T(z-\frac{d}{2})}\right\}, (S22)$$

$$\varepsilon_{xz}^\pm = C\left(1-\frac{\xi_T^2}{2}\right)\sqrt{1-\xi_L^2}\left\{e^{\kappa_L(z-\frac{d}{2})}-e^{\kappa_T(z-\frac{d}{2})}\right\}, (S23)$$

$$\omega_{zx}^\pm = -C\frac{\xi_T^2}{2}\sqrt{1-\xi_L^2}e^{\kappa_T(z-\frac{d}{2})}. (S24)$$

Note that we now omit the operation of taking real parts. Due to the free surface boundary conditions used to derive the solution, $\varepsilon_{xz}$ vanishes at the surface and is smaller by a factor of $kd$ in the magnetic region than $\varepsilon_{xx,zz}$ and $\omega_{zx}$. Thus in the thin film limit, one can ignore $\varepsilon_{xz}$ in $\mathbf{h}_1, \mathbf{h}_3$ and $\mathbf{h}_4$.

The linearized Landau-Lifshitz-Gilbert equation with Gilbert damping $\alpha$, after Fourier transforms in time and space, reads

$$\binom{n_v}{n_z} = \frac{1}{(H+H_v)(H+H_z)-\alpha^2 H_\omega^2+i\alpha H_\omega(2H+H_v+H_z)-(H_\omega-H_{DMI})^2}$$

$$\times \begin{pmatrix} H+H_z+i\alpha H_\omega & -i(H_\omega-H_{DMI}) \\ i(H_\omega-H_{DMI}) & H+H_v+i\alpha H_\omega \end{pmatrix}\binom{h_v}{h_z}, \quad (S25)$$

where we set $\theta = \phi$ for simplicity (i.e. ignoring the effect of cubic anisotropy on the ground state), introduced $H_\omega = \omega/\gamma\mu_0$, and the components of the total effective field $\boldsymbol{h}^\pm$ from SAW with wavenumber $\pm k$ are given by

$$h_v^\pm = i\frac{C\cos\phi}{\mu_0 M_s}\frac{\xi_T^2}{2}\left\{\pm b_1(2-\xi_T^2)\sin\phi + \frac{\hbar\omega S}{V}\sqrt{1-\xi_L^2}\right\}, \quad (S26)$$

$$h_z^\pm = \frac{C\cos\phi}{\mu_0 M_s}\left\{\left(K_\perp - \frac{\mu_0 M_s^2}{2} - K_\parallel - K_c\cos^2\phi\right)\xi_T^2\sqrt{1-\xi_L^2} - Dk\xi_L^2\left(1-\frac{\xi_T^2}{2}\right)\right\}. \quad (S27)$$

The constant $C$ is irrelevant as it multiplies the overall magnitude of the excited spin waves. The energy absorbed from SAW into spin waves per unit time is expected to be proportional to the power $P_{\pm k}$ dissipated by the spin waves, which is equal to the amount fo work done by $\boldsymbol{h}^\pm$

$$P = \mu_0 M_s \left\langle \text{Re}\left[\frac{dn}{dt}\right]\cdot \text{Re}[h^\pm]\right\rangle, \quad (S28)$$

where the angled bracket denotes averaging over one period of SAW. Substituting Eqs. , (S25), (S26) and (S27), one derives the formula for SAW absorption

$$P_{\pm k} = \frac{C^2\alpha\omega^2\xi_T^4\cos^2\phi/2|\gamma|\mu_0 M_s}{\{(H_\omega - H_{DMI})^2 - (H+H_v)(H+H_z) + \alpha^2 H_\omega^2\}^2 + \alpha^2 H_\omega^2(2H+H_v+H_z)^2}$$
$$\times [\{(H+H_z)^2 + (H_\omega - H_{DMI})^2 + \alpha^2 H_\omega^2\}(\pm\rho_{ME}\sin\phi + \rho_{SR})^2]$$
$$+ 2(H_\omega - H_{DMI})(2H+H_v+H_z)(\pm\rho_{ME}\sin\phi + \rho_{SR})\rho_{MR}$$
$$+ \{(H+H_v)^2 + (H_\omega - H_{DMI})^2 + \alpha^2\Omega^2\}\rho_{MR}^2, \quad (S29)$$

The parameters $\rho_{ME,SR,MR}$ measure, in unit of energy density, the contributions from magneto-elastic, spin-rotation and magneto-rotation couplings respectively, defined by

$$\rho_{ME} = b_1\left(1-\frac{\xi_T^2}{2}\right), \rho_{SR} = \frac{\hbar\omega S}{2V}\sqrt{1-\xi_L^2}, \quad (S30)$$

$$\rho_{MR} = \left(K_\perp - \frac{\mu_0 M_s^2}{2} - K_\parallel - K_c\cos^2\phi\right)\sqrt{1-\xi_L^2} - \frac{Dk\xi_L^2}{2\xi_T^2}\left(1-\frac{\xi_T^2}{2}\right), \quad (S31)$$

It can be clearly observed that nonreciprocity, i.e. dependence of $P_{\pm k}$ on the sign $\pm$, comes from cross terms between $\rho_{ME}$ and $\rho_{SR}$ or $\rho_{ME}$ and $\rho_{MR}$. Any $\pm$ dependence is accompanied by a factor $\sin\phi$, which is dictated by the time-reversal symmetry. Note that in Eq. (S29), the dependence of the absorption on the strength of external magnetic field has been all explicitly spelled out so that it may be directly used to fit the SAW transmission data.

**Section S2. Details of the attenuation fitting**

Although the expression ($S29$) can in principle be directly compared with the measured attenuation of SAWs, fitting of the data in practice is not entirely straightforward due to the relatively large number of fitting parameters. It should also be mentioned that the modeling is oversimplified in some aspects so that some discrepancies between the data and the theory are inevitable due to the neglected factors including crystalline anisotropy of $LiNbO_3$, roughness of the interfaces, and the difference in elastic properties between the different materials among other things. To extract the salient features of the relevant physics under discussion in this article, we split the fitting procedure into several steps, which is guided by the underlying physics. Throughout, fitting functions are denoted by block capitals and the corresponding script letters are used for the experimental data to be fitted.

**Step 1: Lorentzian fitting.**

Let $\mathcal{P}_{\pm k}(H, \phi)$ be the attenuation data as a function of the external field $H$ for a fixed angle $\phi$ and wavenumber $\pm k$. We fit each data set $\mathcal{P}_{\pm k}(H, \phi)$ by a function $P_{\pm k}(\phi)$ with three parameters $\mathcal{A}_{\pm k}(\phi), \mathcal{H}^{res}_{\pm k}(\phi), \Delta \mathcal{H}_{\pm k}(\phi)$:

$$P_{\pm k}(\phi) = \frac{1}{2\pi} \frac{\mathcal{A}_{\pm k}(\phi)\, \Delta \mathcal{H}_{\pm k}(\phi)}{\{H - \mathcal{H}^{res}_{\pm k}(\phi)\}^2 + \Delta \mathcal{H}_{\pm k}(\phi)^2/4} \quad (S32)$$

The best fit values for the parameters are now fed into the next steps as the data points.

**Step 2: Resonance field fitting.**

We determine five parameters $A, K_\perp, K_\parallel, K_c, D$ by fitting the data $\mathcal{H}^{res}_{\pm k}(\phi)$ as a function of $\phi$ by the function

$$H^{res}_{\pm k}(\phi) = \frac{-H_v(\phi) - H_z(\phi) + \sqrt{\{H_v(\phi) - H_z(\phi)\}^2 + 4\{\omega/\gamma\mu_0 - H_{DMI}(\phi)\}^2}}{2} \quad (S33)$$

where $H_{v,z,\text{DMI}}(\phi)$ are given by Eqs. ($S7$) - ($S9$). We use the value $|\gamma| = 29.4$ quoted from Ref. and measure $M_s$ independently by MPMS. The best fit values for the parameters are given in TABLE. S1 and plotted in Fig. S1. The DMI constant value presented in the main text was obtained at this step.

**Table S1. Summary of resonance field fitting.**

|  | $A(J/m^2)$ | $K_\perp(J/m^3)$ | $K_\parallel(J/m^3)$ | $K_c(J/m^3)$ | $D(J/m^2)$ |
|---|---|---|---|---|---|
| Best fit value | $9.138 \times 10^{-11}$ | $6.134 \times 10^5$ | $7.212 \times 10^3$ | $-9.023 \times 10^3$ | $8.9 \times 10^{-5}$ |
| Fitting error | $1.112 \times 10^{-12}$ | $2.059 \times 10^{-15}$ | $6.30 \times 10^{-16}$ | $9.341 \times 10^{-16}$ | $1.1 \times 10^{-5}$ |

Note that the value of $A$ is presumably overestimated due to the unquantifiable spatial variation across the film thickness costing significant exchange energy.

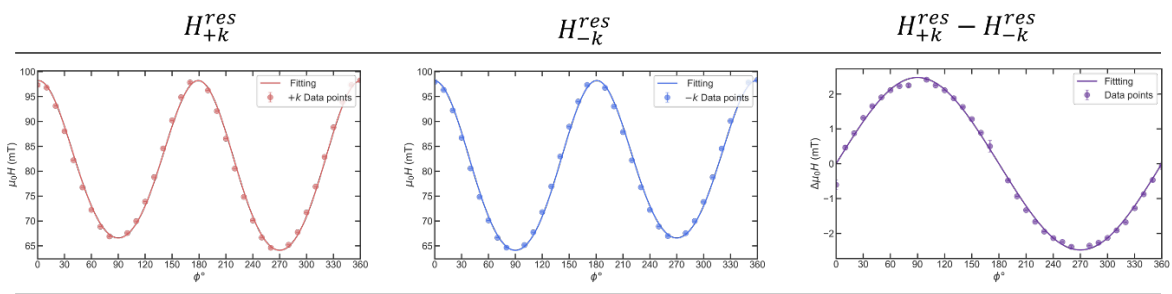

Fig. S1. Fittings for $\mathcal{H}^{res}_{+k}(\phi)$, $\mathcal{H}^{res}_{-k}(\phi)$ and $\mathcal{H}^{res}_{+k}(\phi) - \mathcal{H}^{res}_{-k}(\phi)$, respectively

**Step 3: Linewidth fitting.**

In our simple phenomenology, the linewidth is independent of the angle and given by $\Delta H = \alpha|\omega/2\gamma\mu_0|$. We determine the value of Gilbert damping constant $\alpha$ by equating $\Delta H$ to the $\phi$- and $\pm k$-average of $\mathcal{H}_{\pm k}(\phi)$. The estimated value is $\alpha = 0.059873$ and the fitting results are plorted in Fig. S2.

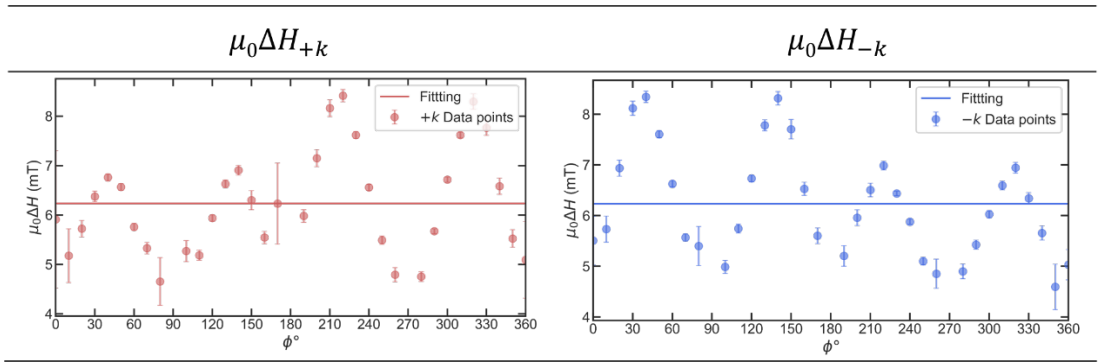

Fig. S2. Fitting for $\Delta H_{+k}$ and $\Delta H_{+k}$ respectively.

**Step 4: Amplitude fitting.**

Finally, we carry out a fitting of the amplitude data $\mathcal{A}_{\pm k}(\phi)$ by the function

$$\mathcal{A}_{\pm k}(\phi) = \frac{\tilde{C}^2 \cos^2 \phi}{\{2H^{res}_{\pm k}(\phi) + H_v(\phi) + H_z(\phi)\}^2}$$

$$\times \left[ \{H^{res}_{\pm k}(\phi) + H_z(\phi)\}^2 + \left\{\frac{\omega}{\gamma \mu_0} - H_{DMI}(\phi)\right\}^2 \right] (r_2 \pm \sin \phi)^2$$

$$+ 2 \left\{\frac{\omega}{\gamma \mu_0} - H_{DMI}(\phi)\right\} \{2H^{res}_{\pm k}(\phi) + H_v(\phi) + H_z(\phi)\} (r_2$$

$$\pm \sin \phi) r_1 + \{H^{res}_{\pm k}(\phi) + H_v(\phi)\}^2 + \left\{\frac{\omega}{\gamma \mu_0} - H_{DMI}(\phi)\right\}^2 r_1^2 \quad (S34)$$

The functions and parameters appearing in the above equation have all been obtained in the previous steps except for those to be fitted, i.e. $\tilde{C}$, $r_1$ and $r_2$. The overall constant $\tilde{C}$ does not contain any meaningful information. The other two $r_{1,2}$ are the central objects of interest in this study, which respectively measure the ratio of the magneto-rotation and spin-rotation coupling energies $\rho_{MR}$, $\rho_{SR}$ to the magentoelastic coupling energy $\rho_{ME}$: $r_1 = \rho_{MR}/\rho_{ME}$, $r_2 = \rho_{SR}/\rho_{ME}$ (c.f. Eqs. ($S30$) and ($S31$)). The nonreciprocity arises from the terms that are linear in $r_{1,2}$. It turns out that these two parameters are highly degenerate: both of them can fit the data equally well on their own and when being fitted at the same time, the error bars tend to be much greater than when only one of them is fitted. The results of the fitting are given in TABLE. S2 and plotted in Fig. S3. In the end, we convert $A_{\pm k}(\phi)$ into $P_{\pm k}(\phi)$ and plot $P_{\pm k}(\phi)$ and rectifier ratio $[P_{+k}(\phi) - P_{+k}(\phi)]/[P_{+k}(\phi) + P_{+k}(\phi)]$ in Fig. S4.

**Table S2. Summary of amplitude fitting.**

|  | $C$ | $C_{err}$ | $r_1$ | $r_{1err}$ | $r_2$ | $r_{2err}$ |
|---|---|---|---|---|---|---|
| Fitting with $r_2 = 0$ | 4.146436 | 0.041575 | 0.179734 | 0.013949 | N/A | N/A |
| Fitting with $r_1 = 0$ | 4.14431 | 0.041322 | N/A | N/A | -0.08191 | 0.006291 |
| 3-parameter fitting | 4.098618 | 0.039964 | -4.278479 | 1.154548 | -2.025804 | 0.524107 |

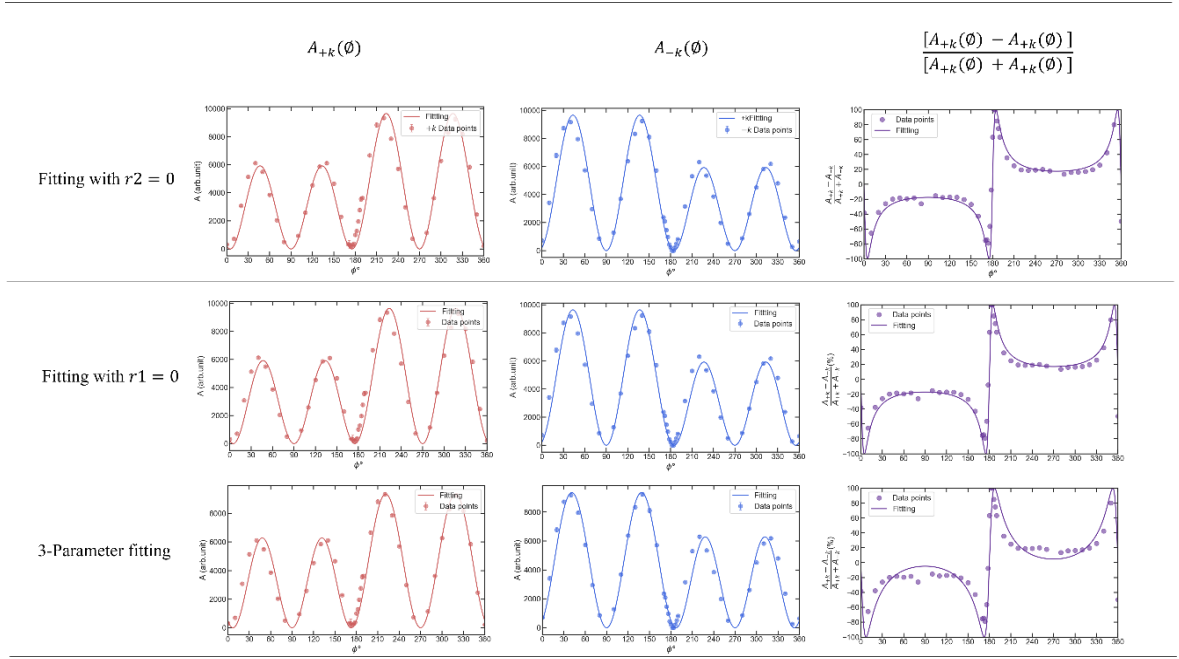

Fig. S3. Fitting for $A_{+k}(\phi)$, $A_{+k}(\phi)$ and $[A_{+k}(\phi) - A_{+k}(\phi)]/[A_{+k}(\phi) + A_{+k}(\phi)]$ under conditions $r2 = 0$, $r1 = 0$ and $r1, r2 \neq 0$, respectively.

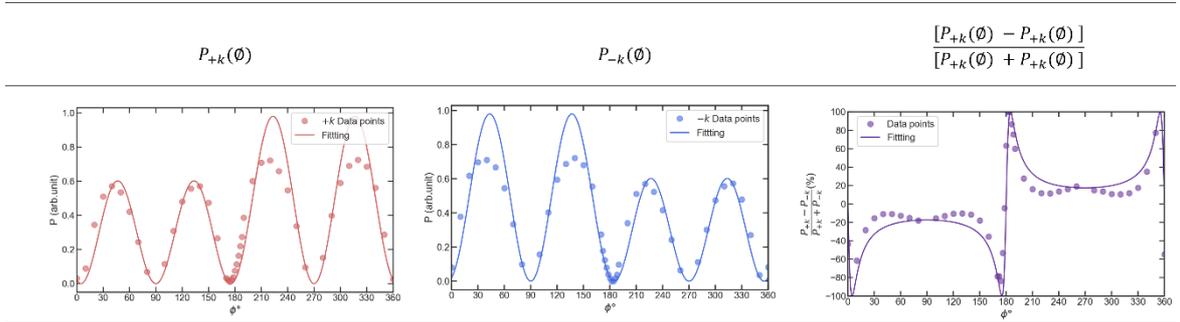

Fig. S4. Fitting for $P_{+k}(\phi)$, $P_{+k}(\phi)$ and $[P_{+k}(\phi) - P_{+k}(\phi)]/[P_{+k}(\phi) + P_{+k}(\phi)]$ under the condition $r2 = 0$.

Even though the nonreciprocity data alone is insufficient to decide which of the magneto-rotation and spin-rotation couplings is the dominant mechanism, we can argue in favor of the former by considering how plausible the best fit values of $r_{1,2}$ are. First of all, we note that the value of $\rho_{MR}$ is completely known from the resonance field fitting and estimated to be $\rho_{MR} \sim -10^6 [\text{J}/\text{m}^3]$. In order to estimate $\rho_{SR}$, one would need to know the effective spin density $S/V$ for CoFeB thin films. Although it cannot be precisely determined due to the uncertainties in the microscopic magnetic structure, one could safely assume $S/V < 10^{30} [\text{m}^3]$ since $S \sim$ $O(1)$ and the unit cell size cannot be smaller than $1[\text{Å}]$. Thus for $\omega = 2\pi \times 6.1[\text{GHz}]$, one obtains $\rho_{SR} < 6\pi \times 10^5[\text{J}/\text{m}^3]$. Therefore, $|r_2|$ would be at best comparable to $|r_1|$ even in the most optimistic scenario. While we do not know the value of $b_1$ for our sample, typical values for transition metals are of order $10^7[\text{J}/\text{m}^3]$ (23) so that the best fit value $r_1 \approx 0.2$ is very reasonable while achieving $r_2 \approx -0.1$ would require a significantly lower magnetostriction for CoFeB than Co or Fe alone. These estimates also suggest that the shear strain mechanism should be far less effective than the magneto-rotation coupling for our thin films as $b_2$ should be of the same order as $b_1$ and $kd \sim 1/500$. Therefore, we conclude that the observed giant nonreciprocity is mainly due to the magneto-rotation coupling induced by the uniaxial anisotropy field $K_u = K_\perp - \mu_0 M_s^2/2$. Although we are unable to exclude a contribution from the spin-rotation coupling, it would be at most of similar order of magnitude to the contribution from the magento-rotation coupling.

**Section S3. Influence of anisotropic substrate on angular dependence attenuation**

The power dissipated by spin waves is given by

$$P = -\frac{\alpha M_s \omega^2}{2\gamma} [\{(H + H_v)(H + H_z) - (H_\omega - H_{\text{DMI}})^2\}^2 + \alpha^2(2H + H_v + H_z)^2 H_\omega^2]^{-1}$$
$$\times [\{(H + H_z)^2 + (H_\omega - H_{\text{DMI}})^2\}|h_v|^2$$
$$+ \{(H + H_v)^2 + (H_\omega - H_{\text{DMI}})^2\}|h_z|^2 + 2(2H + H_v + H_z)(H_\omega$$
$$- H_{\text{DMI}})\Im[\overline{h_z}h_v]]$$

$$= \frac{\alpha M_s \omega^2}{2\gamma} [\{(H + H_v)(H + H_z) - (H_\omega - H_{\text{DMI}})^2\}^2 + \alpha^2(2H + H_v + H_z)^2 H_\omega^2]^{-1}$$
$$\times \{|(H + H_z)h_v + i(H_\omega - H_{\text{DMI}})h_z|^2$$
$$+ |(H + H_v)h_z - i(H_\omega - H_{\text{DMI}})h_v|^2\} \quad , (S35)$$

This is essentially Eq. (29) of the supplemental, written in terms of $h_{v,z}$ instead of $\rho_{\text{ME,MR,SR}}$ via Eqs. (26) and (27). Assuming near resonance $H \sim H_{\text{res}}$, we separate it into the Lorentzian and the residual amplitude

$$P \approx \frac{\alpha |H_\omega| A/\pi}{(H - H^{\text{res}})^2 + \alpha^2 H_\omega^2}, (S36)$$

$$A = \frac{\pi}{2} \frac{\mu_0 M_s |\omega|}{(2H^{\text{res}} + H_v + H_z)^2}$$
$$\times \{|(H + H_z)h_v + i(H_\omega - H_{\text{DMI}})h_z|^2$$
$$+ |(H + H_v)h_z - i(H_\omega - H_{\text{DMI}})h_v|^2\}, (S37)$$

We are doing this splitting because this model evidently fails to fit the observed anisotropic linewidth data (which is addressed later in theses notes) so that the amplitude part should be isolated in comparing the theory with the data. If the magnetic resonance is isotropic, i.e. $H_z = H_v$, one also has $(H + H_v)^2 = (H + H_z)^2 = (H_\omega - H_{\text{DMI}})^2$ at the resonance. Although this approximation is not very good in the present setup where the dipolar shape anisotropy is clearly visible, for simplicity we take it here. Conventionally choosing $H_\omega > 0$, one obtains at the resonance

$$A = \frac{\pi}{4} \mu_0 M_s |\omega| |h_v + i h_z|^2, (S38)$$

With the cubic magneto-elastic coupling and out-of-plane uniaxial anisotropy, the effective magnetic field generated by acoustic waves is given by

$$h_v = \frac{1}{\mu_0 M_s} \{b_1(\epsilon_{xx} - \epsilon_{yy}) \sin 2\phi - 2b_2 \epsilon_{xy} \cos 2\phi\}, (S39)$$

$$h_z = -\frac{2}{\mu_0 M_s} \{(b_2 \epsilon_{zx} + K_\perp \omega_{zx}) \cos \phi + (b_2 \epsilon_{zy} + K_\perp \omega_{zy}) \sin \phi\}, (S40)$$

Suppose that the surface acoustic wave propagates in the $x$ direction, but still has a nonzero $y$ component of the deformation. In our original analysis, we did not include this component since it is absent for SAWs in isotropic media. The boundary conditions force $\epsilon_{zx} = \epsilon_{zy}$ at the boundary, and the effective magnetic field reduces to

$$h_v = \frac{1}{\mu_0 M_s} \{b_1 \epsilon_{xx} \sin 2\phi - 2b_2 \epsilon_{xy} \cos 2\phi\}, (S41)$$

$$h_z = -\frac{2K_\perp}{\mu_0 M_s} (\omega_{zx} \cos \phi + \omega_{zy} \sin \phi), (S42)$$

We cannot derive analytical expressions for the strain and vorticity tensor components in general anisotropic media, but here the purpose is to capture the qualitative trend. First of all, let us assume $\epsilon_{xx}, \omega_{zx}$ are given by those of the SAWs in isotropic media, meaning they are of a similar order of magnitude and have a phase difference of $\operatorname{sgn}(k)\pi/2$. Next, $\epsilon_{xy} = \partial_x u_y/2, \omega_{zy} = \partial_z u_y/2$ arise from the anisotropy correction so that they are expected to be smaller than $\epsilon_{xx}, \omega_{zx}$. For surface localised waves, one expects $\partial_x \sim ik, \partial_z \sim \kappa > 0, k\kappa \in R$ so that it is reasonable to assume the relative phase between $\epsilon_{xy}$ and $\omega_{zx}$ is also $\pm\pi/2$. Hence we introduce the following parameterisation:

$$b_1 \epsilon_{xx} = s, \quad 2K_\perp \omega_{zx} = ib, \quad 2b_2 \epsilon_{xy} = ce^{i\delta}, \quad 2K_\perp \omega_{zy} = ide^{i\delta}, \quad a, b, c, d, \delta \in R, (S43)$$

The experimental data already suggested $|b/a| \sim 0.35$ and $c, d$ represent the anisotropy correction so that $|c|, |d| \ll |a|$. $a$ and $b$ are even and odd with respect to $k$ respectively, while the behaviour under the sign change of $k$ is not known for $c, d$. However, given $\epsilon_{xy} \sim iku_y/2, \omega_{zy} \sim \kappa u_y/2$, it is expected that one is odd and the other is even. One obtains

$$A = \frac{\pi}{4}\frac{|\omega|}{\mu_0 M_s}\left|a\sin 2\phi + b\cos\phi - (c\cos 2\phi - d\sin\phi)e^{i\delta}\right|^2, (S44)$$

Where cross product of a sin2phi and cos phi give the main nonreciprocal term in amplitude, while cross product of $a\sin 2\phi$ and $c\cos 2\phi$ may account for the additional minor oscillation in the angular dependence of nonreciprocity ratio. These give simple understandings of the angular dependence observed.

**Section S4. Details of 100% non-reciprocity**

In the angular dependence spectrum near $\phi = 180$, there is an abrupt change of nonreciprocity (Fig. 3.). Based the theory, it is expected to achieve significantly high nonreciprocity ratio in this region. Defining non-reciprocity by $(A_+ - A_-)/(A_+ + A_-)$ where $A_\pm$ corresponds to $A$ evaluated for $\pm k$, i.e $\pm b$ respectively, it is expected that 100% nonreciprocity may be achieved at angles where either $A_+$ or $A_-$ is equal to zero. By considering the isotropic case for Eq. (44), with $c = d = 0$, this angle can be determined by the condition

$$2a\sin\phi + b = 0, (S45)$$

since $|b/2a| < 0$, this always has a solution near $\phi = 0$ and if $b > 0$ one gets $A_+ = 0$ at a $\phi < 0$. And obviously $A_+ = 0$ implies $(A_+ - A_-)/(A_+ + A_-) = -1$, i.e. 100% non-reciprocity.

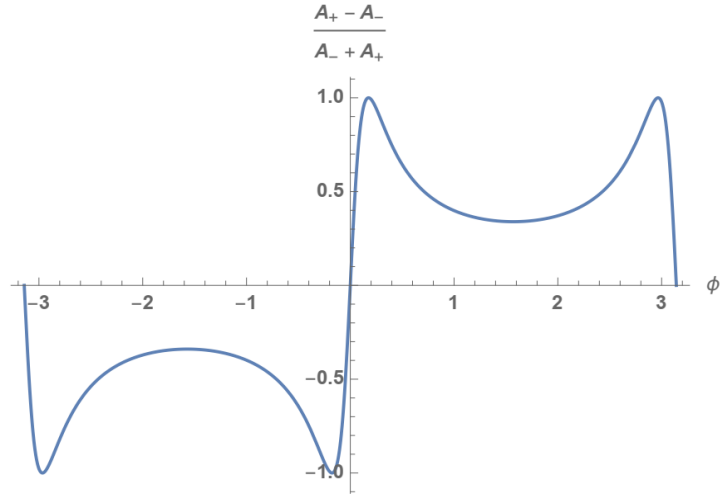

Fig. S5. The relative non-reciprocity of the absorption amplitude *A* when the SAW is assumed isotropic, i.e. $c = d = 0$. We set $a = 1, b = \pm 0.35$. The non-reciprocity reaches 100 % at an angle very close to ϕ = 0.

In experiment, we rotated the magnetic field angle $from\ \phi = 172$ to $\phi = 188$, scanning the nonreciprocity. From the spectra, we confirmed the rapid change of nonreciprocity amplitude and sign. Also, interestingly, when $\phi = 184$, we observed a total flat line for SAW(-k) while maintaining SAW(+k) with robust peak, namely 100% nonreciprocity ratio in accordance with the theory.

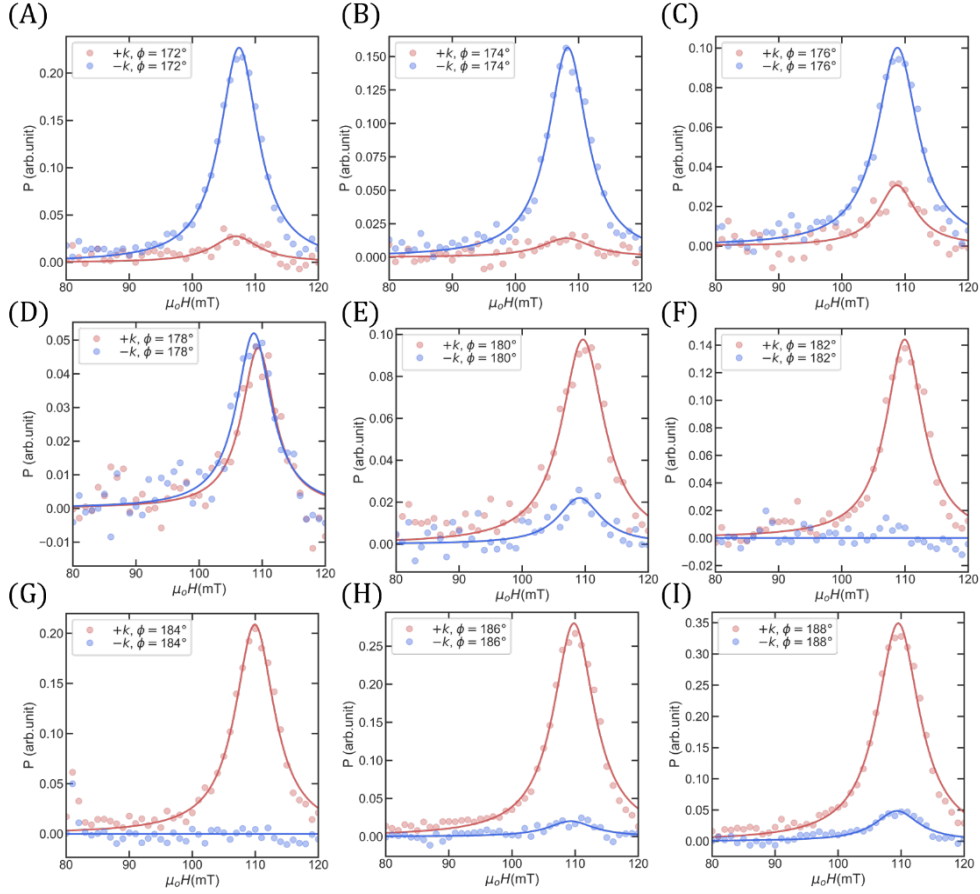

Fig. S6. (A- I) Absorption spectra at $\phi$ =172,174,176,178,180,182,184,186,188, respectively.

**Section S5. Characterization of Dzyaloshinskii-Moriya interaction via Brillioun light scattering spectroscopy**

Dzyaloshinskii-Moriya interaction (DMI) is the antisymmetric exchange coupling, which favors the canting alignment of the neighboring magnetic spins $S_i$ and $S_j$. In recent years, due to its intriguing application in stabilizing magnetic skyrmions and chiral domain walls, DMI has attracted intensive research. In the magnetic heterostructure, DMI appears as a consequence of the broken structural inversion symmetry in the magnet. Among the experimental methods for investigating DMI, Brillouin light scattering (BLS) spectroscopy has been most widely used due to its high sensitivity.

In the presence of the DMI, because of the different canting arrangement, spin waves with wavenumbers $\pm k$ give opposite contributions to the total energy, which results in an asymmetric spin wave dispersion relation. And this asymmetry in $\pm k$ leads to Eq. (S45) (*10, 14, 24, 25*) for estimating DMI constant $D$:

$$\Delta\omega = \frac{\omega(-k) - \omega(k)}{2\pi} = \frac{2Dk|\gamma|}{\pi M_S} \quad (S46)$$

$$k = \text{sgn}(M_x)\frac{4\pi\sin\Theta}{\lambda_{\text{laser}}} \quad (S47)$$

where we take gyromagnetic ratio $|\gamma| = 29.4\text{GHz/T}$ (*26*), saturation magnetization $M_s = 1.5\text{T}$, wavelength of the laser $\lambda_{\text{laser}} = 473\text{nm}$ and $\text{sgn}(M_x)$ is the polarity of the $x$ component of static magnetization, $\Theta$ the angel between incident light and sample plane.

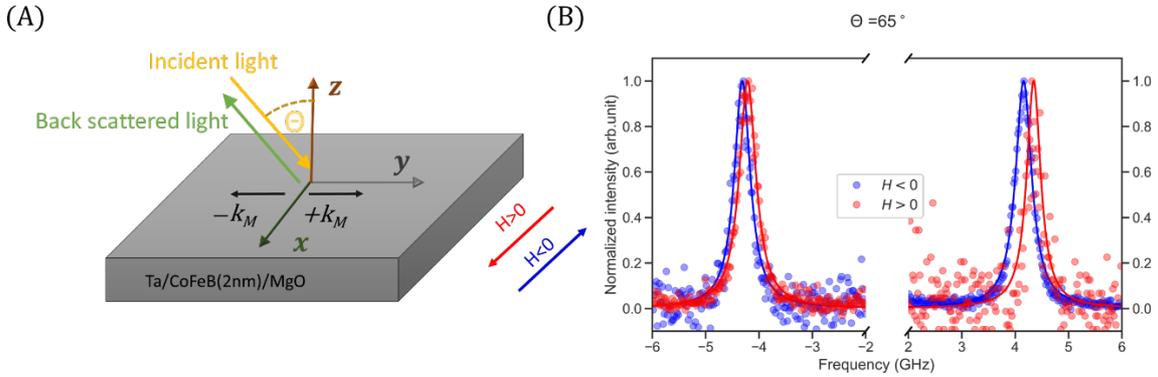

Fig. S7. **Characterization of Dzyaloshinskii-Moriya insteraction via Brillioun light scattering spectroscopy.** **(A)** Schematics of Brillouin light scattering geometry, with scattering plane (in blue), and Cartesian coordinates. **(B)** Brillouin spectra of the Ta/CoFeB(1.6nm)/MgO film measured at incident angle $\Theta = 65°$. Red and blue dots represent spectra measured under applied field H= 50mT along respective $+x$ and $-x$ directions. Solid lines represent Lorentzian fitting of spectra. $k_M$ is the magnitude of wavenumber $k$. Stokes and ant-Stokes peaks were normalized to a peak amplitude of 1, respectively.

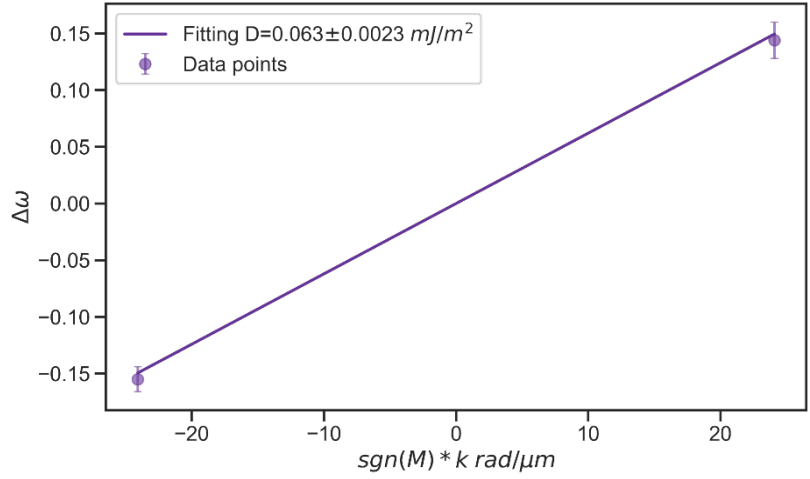

**Fig. S8. Frequency difference *Δω* of ±*k* spin waves as a function of wavevector *k*. Purple circles and solid line denote measured data and fitting by Eq. (*S46*)**

In order to estimate *D*, we performed BLS measurement on the Ta/CoFeB(1.6nm)/MgO thin film in Damon-Eshbach geometry (as depicted in Fig. S5A). Figure Fig. S5B shows measurement of BLS spectra at θ =65° while applying magnetic field *H* at ±50mT. Owing to in-plane momentum conservation of the light scattering process, spin waves traveling with the wavenumber ±*k* appear as anti-Stokes and Stokes peaks, respectively. The difference of spectra center frequency *Δω* in ±*k* are plotted in Fig. S6. By fitting with Eq. (*S46*), we obtain DMI constant $D = 0.063 \pm 0.0023$ mJ/m², which is in a good agreement with the estimation from acoustic ferromagnetic resonance $D_{\text{a-FMR}} = 0.089 \pm 0.011$ mJ/m².